%% file: main.tex
\documentclass[letterpaper, 10 pt, journal, twoside]{IEEEtran}

\IEEEoverridecommandlockouts                            

\usepackage[pdftex]{graphicx}
\graphicspath{{figure/}}
\usepackage{amsmath}
\usepackage{amssymb}
\usepackage{url}

\usepackage{amsthm} 
\usepackage{cite}
\usepackage{color}
\usepackage{comment}
\usepackage{fancyhdr} 
\usepackage{multirow}

\usepackage{hyperref}

\usepackage{algorithm}
\usepackage{algorithmic}
\usepackage{bm}
\usepackage{bbm}

\let\labelindent\relax
\usepackage{enumitem}
\usepackage{mathtools}

\newtheorem{query}{\textbf{Query}}

\newcommand{\sS}{\mathcal{S}}

\newcommand{\sX}{\mathcal{X}}
\newcommand{\sE}{\mathcal{E}}
\newcommand{\sP}{\mathcal{P}}
\newcommand{\sA}{\mathcal{A}}

\newcommand{\sD}{\mathcal{D}}
\newcommand{\sQ}{\mathcal{Q}}

\newcommand{\sC}{\mathcal{C}}
\newcommand{\sJ}{\mathcal{J}}
\newcommand{\sM}{\mathcal{M}}
\newcommand{\sT}{\mathcal{T}}

 \usepackage[dvipsnames]{xcolor}

\newcommand{\xin}[1]{  \ifthenelse{\boolean{showcomments}}
	{ \textcolor{red}{(Xin:  #1)}} {}  }

\usepackage{booktabs} 
\usepackage{array}  
 \usepackage{longtable}

\newcolumntype{L}[1]{>{\raggedright\arraybackslash}m{#1}}

\graphicspath{{Figures/}}

\newboolean{showcomments}
\setboolean{showcomments}{true}

\allowdisplaybreaks 

\title{X-GridAgent: An LLM-Powered Agentic AI System for Assisting Power Grid Analysis}

\author{Yihan (Logon) Wen, Xin Chen
\thanks{Y. Wen and X. Chen are with the Department of Electrical and Computer Engineering, Texas A\&M University, USA; correspondence email: {xin\_chen@tamu.edu}.}
}

\begin{document}

\maketitle



\input{Abstract}

\input{Introduction}

\input{Overview}

\input{Method}

\input{Experiment}

\input{Conclusion}

\input{Reference}

\end{document}

%% file: Abstract.tex
\begin{abstract}
The growing complexity of power system operations has created an urgent need for intelligent, automated tools to support reliable and efficient grid management. Conventional analysis tools often require significant domain expertise and manual effort, which limits their accessibility and adaptability. To address these challenges, this paper presents \emph{X-GridAgent}, a novel large language model (LLM)-powered agentic AI system designed to automate complex power system analysis through natural language queries. The system integrates domain-specific tools and specialized databases under a three-layer hierarchical architecture comprising planning, coordination, and action layers. 
This architecture offers high flexibility and adaptability to previously unseen
tasks, while providing a modular and extensible framework that can be readily expanded to incorporate new tools, data sources, or analytical capabilities. To further enhance performance, we introduce two novel algorithms: (\romannumeral1) LLM-driven prompt refinement with human feedback, and (\romannumeral2) schema-adaptive hybrid retrieval-augmented generation (RAG) for accurate information retrieval from large-scale structured grid datasets. Experimental evaluations across a variety of user queries and power grid cases demonstrate the effectiveness and reliability of \emph{X-GridAgent} in automating interpretable and rigorous power system analysis.

\end{abstract}


%% file: Introduction.tex
\section{Introduction}
\label{sec:intro}

\IEEEPARstart{M}{odern} power systems are undergoing a profound transformation driven by factors such as the high penetration of renewable energy, the widespread deployment of distributed energy resources \cite{muhtadi2021distributed}, the integration of large data center loads \cite{chen2025electricity}, increasing stakeholder involvement, and ambitious decarbonization goals \cite{chen2024towards}. 
These developments have substantially increased the complexity of power system planning, operation, and analysis in dynamic and interconnected environments.
Moreover, 
conventional power system analysis and decision-making require significant manual effort and deep domain expertise. Engineers and operators 
need to navigate multiple specialized software tools and datasets for computation, simulation, and analysis, which are time-consuming, costly, and labor-intensive. As system complexity continues to grow, this manual and fragmented workflow becomes increasingly unsustainable. It underscores the urgent need for advanced artificial intelligence (AI)-driven tools \cite{11131348,chen2022reinforcement} that can automate routine tasks, streamline analysis, and support reliable and efficient grid operations.


In particular, large language models (LLMs) \cite{zhao2023survey}, such as OpenAI’s GPT \cite{achiam2023gpt} and Google’s Gemini \cite{team2023gemini}, have recently demonstrated remarkable capabilities in reasoning, knowledge integration, and context comprehension and generation across various domains \cite{zeng2023large,hadi2023large}.
These advances in LLMs are driving the rapid evolution of generative AI techniques and present significant potential for supporting power system applications. Recent studies have explored the use of LLMs to enhance power system simulations \cite{11086353}, generate power grid models \cite{deng2025power}, and visualize power networks \cite{10747635}, 
among others. A comprehensive review of LLM applications in power systems is provided in  \cite{11129042}.
However, when applied to automate complex domain-specific tasks, the direct use of LLMs reveals several fundamental limitations. First, most existing LLMs are trained to solve general-purpose problems and thus lack the specialized domain knowledge, engineering expertise, and access to proprietary or confidential data, which are essential for real-world power system applications. Second, although LLMs excel at processing and generating natural language, they are not inherently designed for performing precise numerical computations or addressing mathematically rigorous problems such as power flow analysis, optimization, and control \cite{majumder2024exploring}. These limitations pose significant challenges to deploying LLMs as standalone solutions in critical engineering domains such as power systems.

To overcome these challenges, recent advances have introduced \emph{agentic AI} systems \cite{durante2024agent,10849561}, which
are LLM-based autonomous agents equipped with external tools, domain-specific databases, and enhanced reasoning capabilities.
Unlike standalone LLMs, agentic AI systems can dynamically interact with structured data sources, plan and execute multi-step workflows, and perform specialized computations using domain-specific tools. 
A key technique for these systems is {Retrieval-Augmented Generation (RAG)} \cite{fan2024survey,singh2025agentic}, which allows LLMs to retrieve and incorporate relevant information from local specialized knowledge bases, thus grounding responses in accurate and domain-specific data.
In addition, the emerging open standard {Model Context Protocol (MCP)} \cite{anthropic2025mcp,hou2025model} provides a unified framework for seamless communication between LLM agents and external tools. MCP defines standardized interfaces and coordination mechanisms that enable agents to access, query, and utilize software tools, application programming interfaces (APIs), and data services in a modular and extensible manner. 
The integration of these advanced techniques transforms LLMs from passive text generators into autonomous AI agents capable of reasoning, planning, and acting to solve complex real-world tasks.

In this paper, we present a novel LLM-powered agentic AI system, called \textbf{X-GridAgent}, for automating comprehensive power grid analysis using only natural language. This system 
bridges the gap between high-level language interfaces and complex power grid analysis tasks by integrating RAG, MCP, domain-specific tools, and specialized databases
under a hierarchical reasoning architecture. 
X-GridAgent enables users to perform a wide range of professional power system analyses, including power flow analysis, contingency analysis, optimal power flow (OPF), short-circuit calculation, topology search, and more. All the grid analyses are 
conducted using domain-specific tools to ensure trustworthy and interpretable results, and 
merely through conversational queries, \emph{without} the need for direct interaction with simulation software or programming environments. To achieve this, we develop a series of MCP-based tool servers, each tailored to handle a distinct category of grid analysis tasks (see Section~\ref{sec:serana} for details). 
By combining the advanced reasoning capabilities of LLMs with professional tools and structured knowledge bases, X-GridAgent aims to automate complex power grid analysis and make it 
more accessible, interpretable, and efficient for engineers, researchers, system operators, and other stakeholders. 

\begin{figure}
	\centering
    \includegraphics[width=0.49\textwidth]{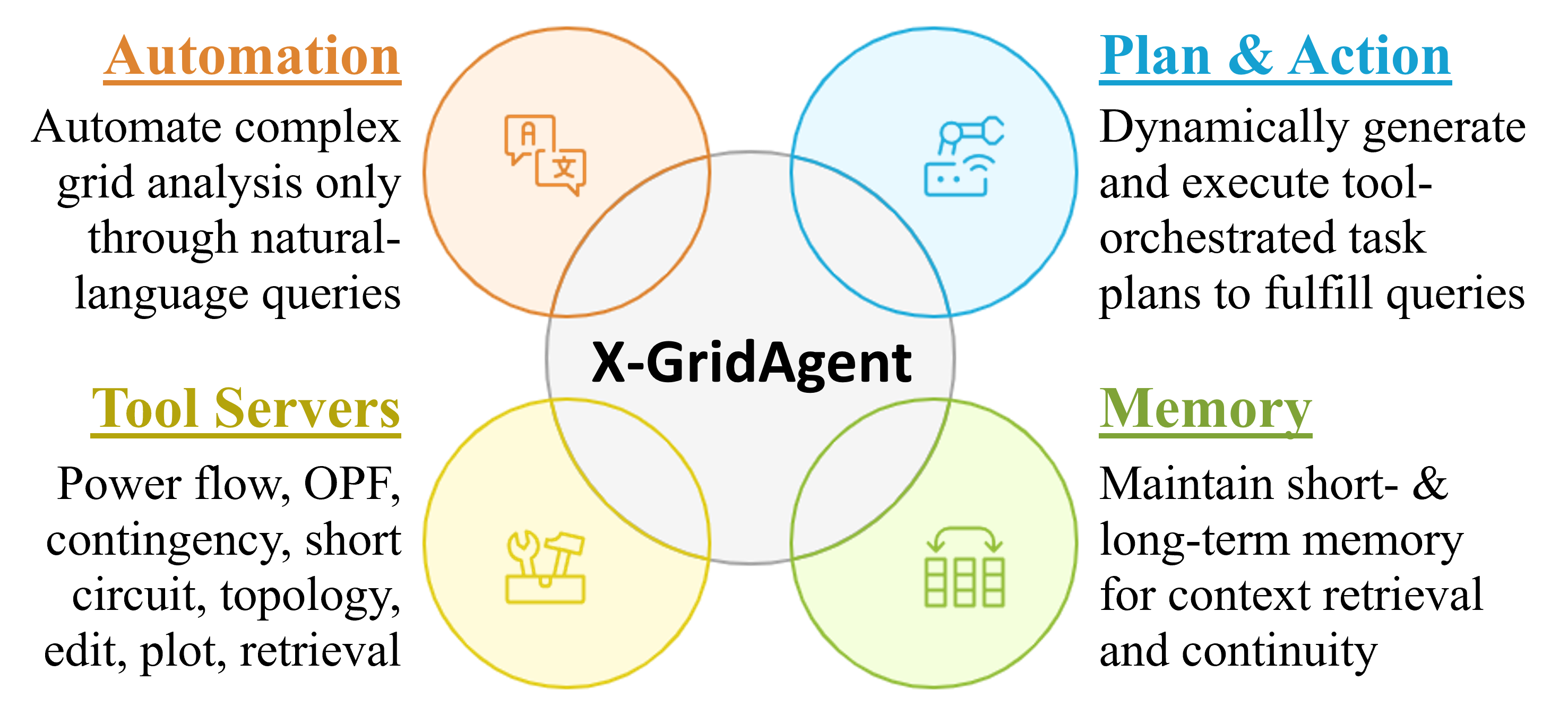}
    \vspace{-10pt}
	\caption{The four key features of the proposed X-GridAgent system.
	}
	\label{fig:feature}
\end{figure}

\begin{figure*}
	\centering
	\includegraphics[width=0.90\textwidth]{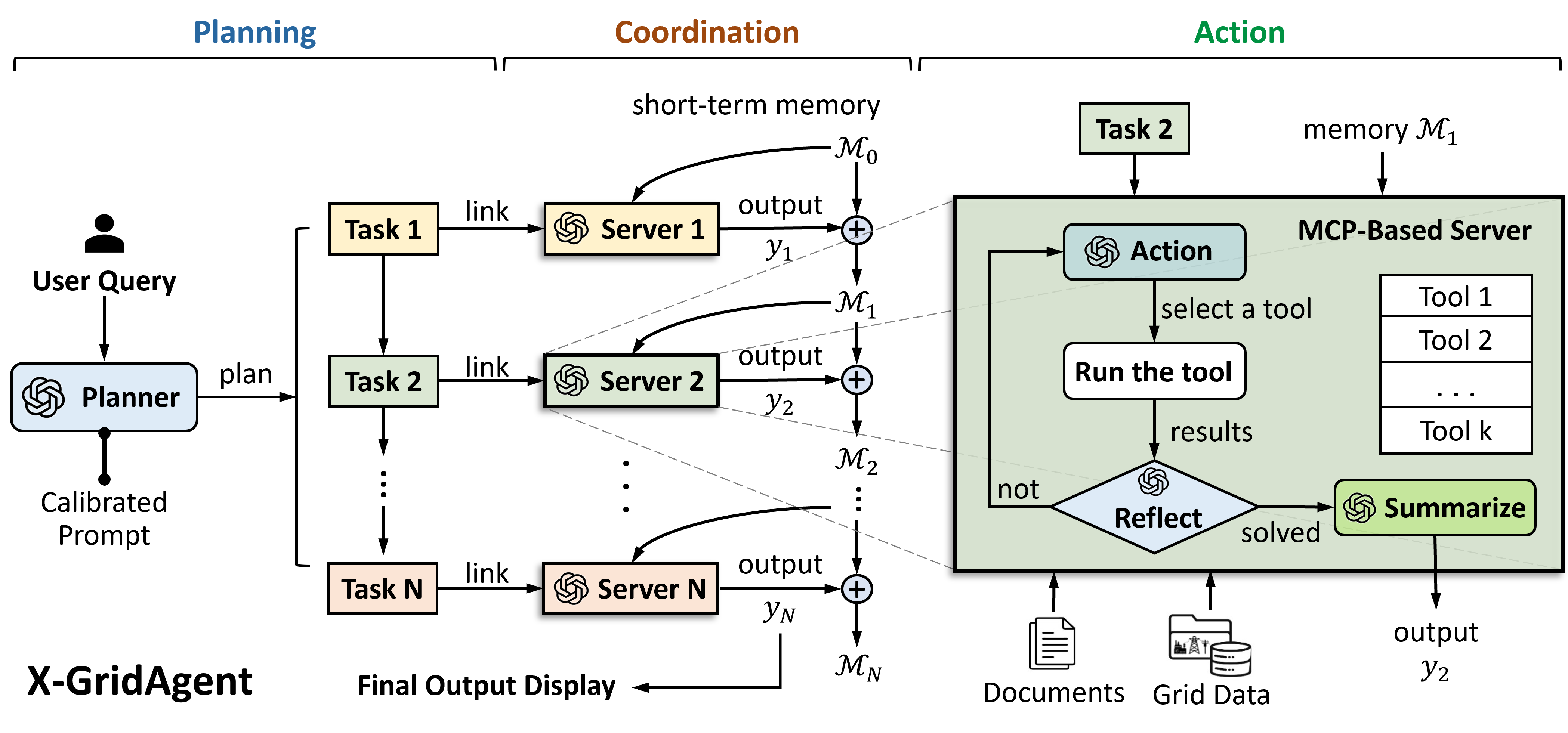}
	\caption{Overview of the three-layer hierarchical architecture of X-GridAgent.
	}
	\label{fig:gridagent}
\end{figure*}

\textbf{\emph{Contributions.}} The key contributions of this work include: 
\begin{enumerate}
    \item We develop an \emph{integrated} agentic AI system \emph{X-GridAgent} to automate complex power grid analysis through natural language queries. Leveraging the reasoning capabilities of LLMs, it 
     dynamically generates task-specific workflows and invokes domain-specific tools for rigorous power system analysis. It also incorporates both short-term and long-term memory for coherent multi-turn interactions and contextual information retrieval from databases. The key features of X-GridAgent are shown in Figure~\ref{fig:feature}.

    \item We design a novel \emph{three-layer hierarchical architecture} for the X-GridAgent system, consisting of the \emph{planning}, \emph{coordination}, and \emph{action} layers, to support complex, multi-step power grid analysis tasks. As shown in Figure \ref{fig:gridagent}, the planning layer interprets user intents and generates a plan of sequential tasks, the coordination layer 
    routes each task to a designated server and manages its execution, and the action layer interfaces with embedded tools to perform computation or operation. This architecture offers high flexibility and adaptability to previously unseen tasks, and provides a \emph{modular} and \emph{extensible} framework that can be easily expanded to incorporate new tools, data sources, or other analytical capabilities.

    \item We propose two novel algorithms to further enhance the performance of X-GridAgent: (\romannumeral1) \emph{LLM-driven prompt refinement with human feedback} and (\romannumeral2) \emph{schema-adaptive hybrid RAG}. The former leverages the LLM to automatically construct and iteratively refine system prompts with minimal input from human experts, which significantly improves efficiency and consistency while substantially reducing the manual effort typically required for prompt engineering. 
 The latter addresses the limitations of conventional RAG methods in retrieving information from large-scale structured power system datasets; it dynamically selects and reconstructs the most relevant data tailored to the user query and
 employs a hybrid retrieval strategy to improve accuracy and reliability.
\end{enumerate}

\vspace{2pt}
\textbf{\emph{Related Work.}} The application of LLM-based agentic AI in the power systems domain has recently garnered growing attention. Prior works have explored a range of use cases, such as bidding strategy generation in electricity markets \cite{11070004}, unit commitment and power dispatch \cite{ren2025can,yang2025large,zhaolarge,zhu2025powercon}, OPF modeling and solving \cite{li2025optdispro,bernier2025powergraph},   power grid control \cite{zhang2025grid}, and fault analysis \cite{saha2025dragent}.
These studies primarily focus on a single type of power system decision-making task, without addressing the full pipeline required for comprehensive and automated power grid analysis. 
Some works generate power system decisions directly from LLMs, without invoking domain-specific tools or validated code-based methods. Such approaches often lack grounding in physical laws and engineering constraints, and therefore cannot guarantee trustworthy or interpretable results. In the context of power grid analysis,
reference \cite{jin2025gridmind} proposes an LLM-based multi-agent AI system that can perform OPF and contingency analysis via natural language queries and function calls. However, this system remains an early-stage prototype with limited functionality and a simple architecture that may not support complex multi-step grid analysis tasks. In \cite{badmus2025powerchain}, an agentic AI system is developed to perform distribution grid analysis, while annotated expert-built
 workflows are incorporated to enhance its generalizability to unseen tasks. In contrast, 
X-GridAgent offers several key advantages over existing agentic AI systems for power grid analysis: (\romannumeral1) it supports comprehensive grid analysis functionality, together with effective memory management and information retrieval capabilities; (\romannumeral2) its hierarchical architecture and modularized server design enable flexible handling of complex and previously unseen tasks, while remaining easily extensible to new data sources and functionalities; and (\romannumeral3) it integrates automatic prompt refinement and customized RAG techniques that significantly enhance its scalability, efficiency, and response quality.
A demonstration video showcasing our X-GridAgent system in action is available online  \cite{gridagent_demo_video}.

The remainder of this paper is organized as follows. Section~\ref{sec:archi} outlines the hierarchical
architecture of X-GridAgent with the three key layers and specialized servers. Section~\ref{sec:method} introduces our  innovative methods on prompt refinement and RAG. Section~\ref{sec:exper} presents the system implementation and test experiments. Conclusions are drawn in Section~\ref{sec:conclusion}.

%% file: Overview.tex
\section{X-GridAgent Architecture and Design}\label{sec:archi}

This section presents the hierarchical architecture of X-GridAgent comprising three fundamental layers: \emph{planning}, \emph{coordination}, and \emph{action}. A series of specialized X-GridAgent servers is designed to perform rigorous power system analysis by integrating with professional tools and external databases.

\subsection{Overview of X-GridAgent Architecture}

The X-GridAgent system features a novel three-layer hierarchical architecture, comprising the \emph{planning}, \emph{coordination}, and \emph{action} layers, which are designed to jointly address complex power grid analysis tasks. The X-GridAgent architecture is illustrated in Figure \ref{fig:gridagent}.
When a natural-language user query $q$ is submitted, the planning layer $\sP$ first interprets the query and generates a workflow plan that decomposes the query into a sequence of tasks to resolve it. Each task is routed to its corresponding domain-specific server for execution. 
Then, the coordination layer $\sC$ manages the sequential execution of tasks and monitors their progress. It also maintains a short-term memory module $\sM$, which stores key intermediate results and relevant contextual information to facilitate effective information exchange across tasks and ensure consistency throughout the workflow. The actual execution of each task is handled by the action layer $\sA$, which selects appropriate tools within the corresponding server to perform professional computations and actions. It also incorporates a reflection mechanism to determine whether the task has been successfully completed. 
Note that the X-GridAgent workflow is not fixed but is dynamically generated based on the specific user query $q$. This three-layer hierarchical architecture enhances the flexibility, reliability, and scalability for automating power system analysis. 
Below, we elaborate on the domain-specific servers in X-GridAgent and the three integrated layers ($\sP, \sC, \sA$).

\subsection{X-GridAgent Servers}\label{sec:serana}

To endow X-GridAgent with rigorous power system analysis capabilities and domain-specific expertise, we develop a series of specialized servers that integrate professional computational tools, structured databases, and technical documentation. 
This integration is implemented using MCP \cite{anthropic2025mcp,hou2025model}, an emerging open standard that provides a unified framework for seamless communication between LLM agents and external tools or data sources. Specifically, tools refer to callable external functions or APIs that AI agents can invoke to retrieve information, perform computations, or execute actions. These tool-integrated servers constitute X-GridAgent’s core capabilities.

In particular,
built on the open-source power system software \emph{Pandapower} \cite{pandapower2018}, we have implemented eight modular MCP-based servers, each tailored to handle a distinct category of domain-specific tasks and computational problems. These X-GridAgent servers include (1) the \texttt{Retrieval} server, which retrieves relevant information from embedded documentation and power grid datasets using advanced RAG techniques; (2) the \texttt{PowerFlow} server, which performs AC and DC power flow studies; (3) the \texttt{OPF} server, which solves OPF problems to determine optimal generation dispatch and costs under network constraints; (4) the \texttt{Contingency} server, which conducts N-1 contingency analysis on power lines or  transformers and identifies violations of operational limits; (5) the \texttt{ShortCircuit} server, which computes short-circuit currents in accordance with the IEC 60909 standard \cite{DIN_EN_60909_0} under common fault types; 
(6) the \texttt{Topology} server, which supports connectivity assessment, island detection, and network traversal analysis; (7) the \texttt{Edit} server, which modifies grid parameters and configurations (e.g., load values, voltage limits, and line addition or removal); and (8) the \texttt{Plot} server, which visualizes power networks and plots related datasets.

Each server described above is associated with a set of related tools (or functions), enabling flexible execution and extended  functionality. These tools can be dynamically invoked accroding to the user query. For example, in the \texttt{PowerFlow} server, the functions  \emph{solve\_AC\_powerflow()} and  \emph{solve\_DC\_powerflow()} are available to perform AC and DC power flow calculations, respectively. Users can also specify input parameters for these functions directly through natural-language queries. For instance, a user may request a specific solution algorithm, such as Newton-Raphson or fast-decoupled methods, by simply stating this preference in the query; it is then automatically translated into appropriate input parameters for tool invocation.
In addition, two default functions, \emph{NetInit()} and  \emph{NetSave()}, are included in all servers to support the loading and saving of power network cases for analysis.

By decomposing a user query into a chain of structured tasks and routing each task to a specialized server for execution, X-GridAgent can flexibly and effectively handle complex power grid analysis queries. 
Moreover, our \emph{modularized} server framework is inherently extensible, as new capabilities can be integrated easily by deploying additional servers or augmenting existing ones with more tools (functions). This design enables X-GridAgent to continuously evolve in response to emerging analytical requirements and advancements in domain-specific tools, allowing it to remain adaptable, up-to-date, and flexible for supporting a wide range of power system applications. 


\subsection{Planning Layer } \label{sec:planning}

Given a user query $q$ as input,
the planning layer $\sP$ is designed to interpret its intent and generate an executable step-by-step plan composed of logically structured tasks to address the query. Specifically, the planning layer $\sP$ combines the user query $q$ with a well-calibrated system prompt $\pi_{\sP}^*$ to form an augmented input, which is then passed to the LLM for reasoning and plan generation. 
See Section \ref{subsec:apo} for a detailed introduction to our prompt design approach. 
Guided by the system prompt for planning $\pi_{\sP}^*$ and leveraging the LLM’s reasoning capabilities, the planning layer $\sP$ decomposes the query $q$ into a chain of tasks $\Gamma$ executed in sequence: 
\begin{align}
    \Gamma \coloneqq \{ \sT_1\to \sT_2\to \cdots \to \sT_N \} = \sP(q;\pi_{\sP}^*),
\end{align}
where each task $\sT_i$ is configured with a specific objective and is linked to a designated X-GridAgent server for execution.

For example, consider an illustrative multi-step user query: 
\begin{query} \label{query:ex1}
   ``\textit{Can you run AC power flow on the IEEE 118-bus grid to find the top 3 most heavily loaded lines, and then run contingency analysis on those lines to see if there are any operational limit violations?}"
\end{query}
In response to Query \ref{query:ex1}, the planning layer $\sP$ generates the following sequence of tasks defined by server-objective pairs:

{
\setlist[itemize]{leftmargin=1em}
\begin{itemize}
    \item {[Task 1]}: $\sT_1= \{${Server}: \texttt{PowerFlow}, {Objective}: \textit{Run AC power flow analysis on the IEEE 118-bus system}$\}.$
    \item {[Task 2]}: $\sT_2= \{${Server}: \texttt{Retrieval}, {Objective}: \textit{Retrieve the top three power lines with the highest loading ratios from the power flow results}$\}.$
      \item {[Task 3]}: $\sT_3= \{${Server}: \texttt{Contingency}, {Objective}: \textit{Run N-1 contingency analysis by individually tripping the three lines identified in the previous step and report any operational limit violations}$\}.$   
\end{itemize}
}

In this way, unstructured natural-language user queries are translated into structured executable workflows represented as sequences of multi-step tasks by the planning layer $\sP$. These workflows are not fixed but are dynamically generated according to specific user queries, enabling flexible, explainable, and automated power grid analysis. 
The chain of tasks $\Gamma$ is then passed to the coordination layer $\sC$ for management and execution, as detailed in the next subsection.

\subsection{Coordination Layer}\label{sec:coordination}

The coordination layer $\sC$ orchestrates the execution of the chain of tasks  
$\Gamma$ generated by the planning layer. It routes each task to the designated X-GridAgent server, manages the inputs and outputs of each step, and maintains short-term memory to track intermediate results and contextual information throughout the execution process. The actual execution of each task within a X-GridAgent server is carried out by the action layer 
$\sA$, which is introduced in Section~\ref{sec:action}.

\emph{Short-Term Memory Management}.
The tasks in the planned sequence $\Gamma$ are logically connected, with the output of each preceding task providing necessary context and numerical results for those that follow. 
To support these inter-task dependencies, the coordination layer maintains a short-term memory 
 module $\sM$, which is incrementally updated after each task and exposes relevant information to downstream tasks. As illustrated in Figure \ref{fig:gridagent}, for each task $\sT_i$  ($i=1,2,\cdots$), the current memory $\sM_{i-1}$ is concatenated with the task to form an augmented input to the designated server for execution. The initial memory $\sM_0$ captures historical conversations and prior results. 
 After task $\sT_i$ completes, its output $y_i$ is incorporated into the short-term memory to produce an updated $\sM_i$, which is then used by the subsequent task $\sT_{i+1}$. This short-term memory management mechanism enables seamless information flow across tasks, maintains an up-to-date record of contextual information, and supports multi-turn interactions with coherent responses to successive user queries. 

\subsection{Action Layer}\label{sec:action}


By employing MCP and leveraging the advanced reasoning capabilities of the LLM, the action layer $\sA$ autonomously selects and invokes the appropriate tools in the designated X-GridAgent server and manages their input-output data flows to fulfill a task. 
As illustrated in Figure~\ref{fig:gridagent},
the workflow within the action layer $\sA$ is iterative and adapts to each task $\sT_i$ ($i=1,2,\cdots$). 
Specifically, the action layer $\sA$ is provided with a list of available tools and their capabilities from the designated X-GridAgent server. 
Based on the specific task context $\sT_i$ and the current short-term memory $\sM_{i-1}$, it selects an appropriate tool for execution and obtains the corresponding results.
A \emph{reflection} mechanism \cite{bo2024reflective} is designed to evaluate whether the task has been successfully solved, leveraging the LLM’s reasoning to assess alignment with the task objective. 
If the task is deemed complete, an LLM-based summarization module generates a concise and relevant output $y_i$. If not, the evaluation outcome, together with prior computational results, is fed back into the reasoning process to guide the selection of another tool for continued execution. This process continues until the cumulative outcomes sufficiently address the current task $\sT_i$ or the maximum number of iterations is reached. The final output $y_i$ for task $\sT_i$ is then concatenated with $\sM_{i-1}$ to produce an updated short-term memory $\sM_i$.

Specifically, in iteration $k$, the above MCP-based tool selection and invocation process defines a mapping:
\begin{align}
    p_k^i = \sA(\sT_i, \sM_i, p_{k-1}^i; \pi_{\sA}^*),  
\end{align}
where $p_k^i$ represents the cumulative outcome at iteration $k$ for task $i$ after a tool invocation. It records prior results and serves as the input context for the next tool invocation, ensuring that reasoning remains stateful and coherent across the tool sequence.
 $\pi_{\sA}^*$ denotes the well-calibrated system prompt for the action layer. Then, the reflection mechanism compares the cumulative outcome $p_k^i$ with the task context $\sT_i$, given by: 
 \begin{align}
     r_k = \text{Reflect}(p_k^i, \sT_i), 
 \end{align}
where $r_k$ represents the reflection result, 
indicating whether the current outcome sufficiently completes the task and identifying any missing information required for completion. The reflection outcome $r_k$ is then used to guide the next reasoning step, influencing subsequent tool selection.

\emph{Illustrative Example}. To illustrate the detailed process, we consider the execution of Task 3 ($\sT_3$) in the \texttt{Contingency} server to address  Query \ref{query:ex1} presented in Section \ref{sec:planning}. After executing Tasks 1 and 2, the top three power lines with the highest loading ratios are identified as line 6, line 7, and line 34. Accordingly, the action layer $\sA$ autonomously invokes a sequence of functions to complete Task 3, as listed below:
{
\setlist[itemize]{leftmargin=1.6em}
\begin{itemize}
    \item [1.] \emph{NetInit}(``analyzednet": ``ieee118")
    \item [2.] \emph{run\_contingency}(``nminus1\_cases": {``line": {``index": [6]}})
    \item [3.] \emph{run\_contingency}(``nminus1\_cases": {``line": {``index": [7]}})
    \item [4.] \emph{run\_contingency}(``nminus1\_cases": {``line": {``index": [34]}})
    \item [5.] \emph{NetSave}()
\end{itemize}
}
It first loads the IEEE 118-bus system from an embedded grid database using the function ``\emph{NetInit()}" and then iteratively invokes the function ``\emph{run\_contingency()}" three times with correct parameter arguments to perform N-1 contingency analysis on the three identified lines and obtain the limit violation results. 
The reflection module assesses whether the task has been successfully completed after each step and consistently reports non-completion as well as the missing results until Step 4 has been executed.  
Lastly, ``\emph{NetSave()}" is invoked by default to save the power network data and computational results.

%% file: Method.tex
\section{Key Technical Innovations}
\label{sec:method}

This section introduces two novel algorithms for enhancing X-GridAgent’s performance: (1) \emph{LLM-driven prompt refinement with human feedback} and (2) \emph{schema-adaptive hybrid RAG}. The former leverages the LLM to automatically construct and iteratively refine system prompts with light input from human experts, guiding X-GridAgent in context-aware reasoning and appropriate tool use. 
The latter addresses the limitations of conventional RAG methods in retrieving information from
large-scale structured power grid datasets
by dynamically selecting the most relevant data tailored to the user query and using a hybrid retrieval
strategy.


\subsection{LLM-Driven Prompt Refinement with Human Feedback}
\label{subsec:apo}

Prompt design \cite{marvin2023prompt} is critical in shaping the reasoning behavior of LLM-based AI agents: it offers contextual grounding, structures the reasoning process, specifies output formats, and communicates available tools and usage guidelines. However, manually crafting system prompts for complex domains such as power grid analysis is labor-intensive and often driven by subjective intuition, leading to inconsistencies and limited generalizability across tasks. Moreover, misalignment between human phrasing and machine interpretation can further degrade response accuracy and efficiency. 
To address these challenges, we propose an \emph{LLM-driven prompt refinement with human feedback} framework for constructing and refining system prompts. The framework leverages the LLM’s capabilities in comprehension, generation, and editing to calibrate prompts by aligning X-GridAgent’s outputs with ground-truth references. Human experts make minor revisions to ensure factual accuracy and correct detail-related errors. This LLM-led, human-in-the-loop prompt refinement process progressively enhances X-GridAgent’s reasoning stability and accuracy. 

\begin{figure*}
	\centering
    \includegraphics[width=0.9\textwidth]{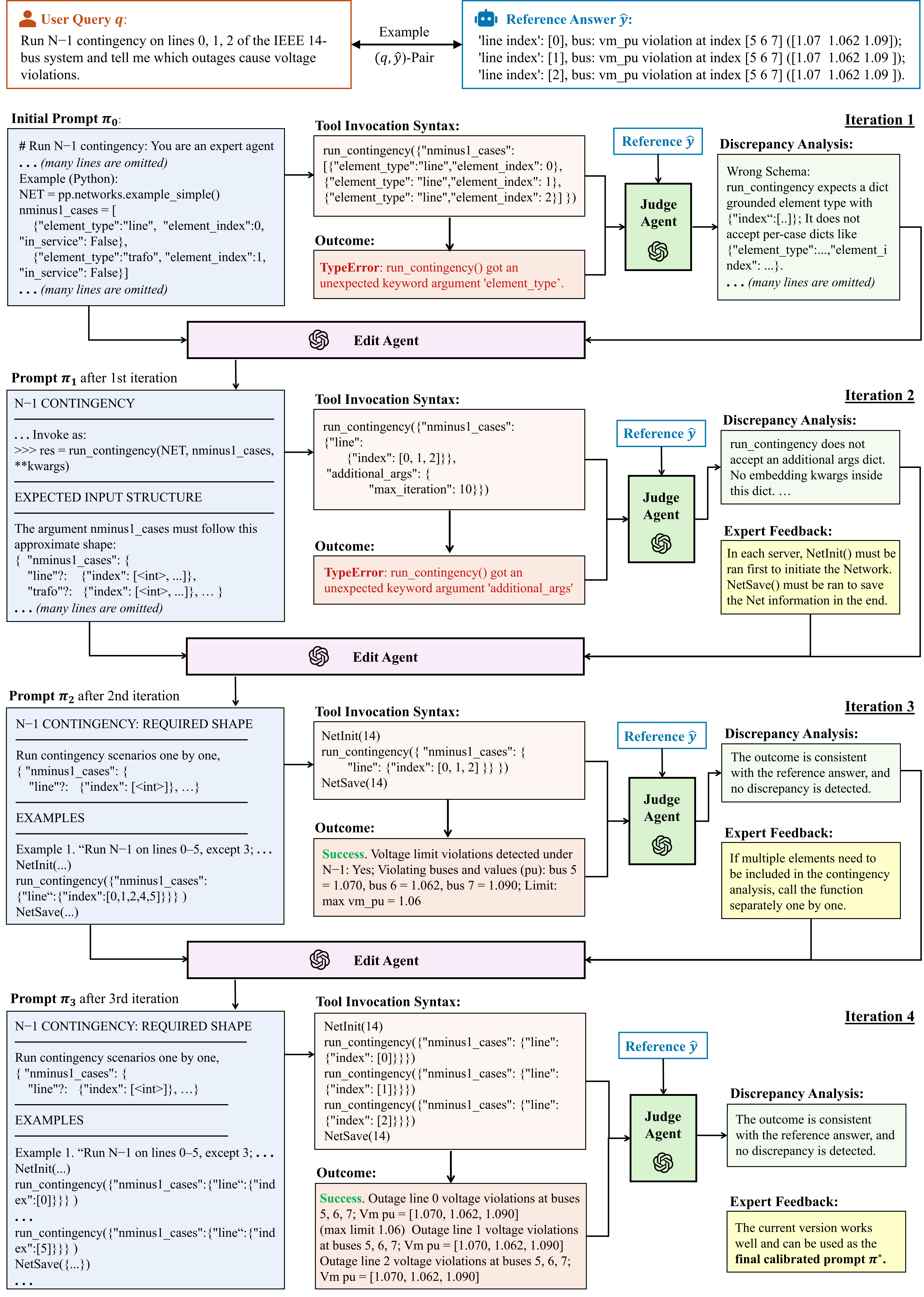}
	\caption{Illustration of the iterative process of the LLM-driven prompt refinement with human
feedback for correctly invoking the ``\emph{run\_contingency()}'' function. (In the third iteration, although the outcome is correct, it does not specify which contingency causes the voltage violation. This oversight is not detected by the judge agent, requiring the human expert to point it out in the feedback.)
	}
	\label{fig:opt}
\end{figure*}

Specifically, let $\pi$ denote the system prompt. Given a user query $q$, the X-GridAgent system configured with prompt $\pi$ produces a final output $y$ as \eqref{eq:ga}: 
\begin{align} \label{eq:ga}
    {y} \;=\; \mathrm{X}\text{-}\mathrm{GridAgent}(q;\pi).
\end{align}
To calibrate the system prompt $\pi$, we use the LLM to automatically generate a set of power system domain-specific queries in natural language, supplemented with some manually authored queries from human experts, to form a training query set $\sQ\coloneqq({q_i})_{i=1}^N$. This set covers a variety of power system analysis tasks, including power flow calculation, contingency analysis, optimal power flow, grid information extraction, and more. 
For each query $q_i$, a corresponding ground-truth reference answer ${\hat{y}_i}$ is produced by human experts using professional tools, which provide verified numerical results and domain-accurate solutions. The resulting dataset $\mathcal{D} \coloneqq \{(q_i, \hat{y}_i)\}_{i=1}^N$ is then used to iteratively refine X-GridAgent's system prompt, starting from an initial version $\pi_0$ and progressing through successive updates $\pi_1 \to \pi_2 \to \cdots \to \pi_k \to \cdots$. 
To support this process, we construct two LLM-based agents: a judge agent $\mathcal{J}$, which automatically evaluates X-GridAgent’s outputs against ground-truth answers, and an edit agent $\mathcal{E}$, which proposes prompt modifications based on the evaluation. 
 
At each iteration $k$, for a selected query $q_k\in\sQ$,  the judge agent $\sJ$ evaluates X-GridAgent's generated output $y_k=\mathrm{X}\text{-}\mathrm{GridAgent}(q_k;\pi_k)$ and produces a structured discrepancy analysis $d_k$ by comparing it with the ground-truth reference answer $\hat{y}_k$:
\begin{align}
    d_k = \sJ(q_k, y_k, \hat{y}_k).
\end{align}
For example, the discrepancy analysis may include feedback such as ``voltage limits are not checked", ``power flow tools are not correctly invoked", and ``generation costs are missing". 
The edit agent $\sE$ then updates the system prompt based on the discrepancy analysis $d_k$:
\begin{align}
    \pi_{k+1} = \sE(\pi_{k},d_k).
\end{align}
This evaluation-edit process can operate autonomously or in a human-in-the-loop mode, where human experts provide feedback by manually revising the discrepancy analysis $d_k$ or directly correcting the updated prompt $\pi_{k+1}$ 
to make necessary adjustments and to better align with domain-specific needs. 

An illustrative example of prompt refinement for correctly invoking the \emph{run\_contingency()} function is shown in Figure~\ref{fig:opt}. The initial system prompt $\pi_0$ is first calibrated by the edit agent $\sE$ to yield $\pi_1$, and then revised to $\pi_2$ with expert feedback. 
The iterative prompt refinement process terminates when the judge agent $\sJ$ and human experts detect no significant discrepancies between $y$ and $\hat{y}$ and marks all queries in $\mathcal{D}$ as passed. The system prompt at this point is taken as the final calibrated prompt $\pi^*$, which is then deployed in X-GridAgent.

\subsection{Schema-Adaptive Hybrid RAG}
\label{subsec:der}

\subsubsection{Background}

RAG \cite{zhao2024dense,ma2023query} is a prominent technique that couples documentation retrieval with text generation to ground LLM responses in 
accurate and domain-specific data sources. 
In a standard RAG pipeline, the documentation is preprocessed into a set of passages or chunks, often with a target length $N$ and optional overlap. Given a user query, a retriever scores and selects the top-$k$ chunks by relevance. 
Based on these selected chunks, the LLM generates an output grounded on the provided evidence, which mitigates hallucinations and improves performance on domain-specific tasks. 

The retrieval mechanisms for RAG have two main types: \emph{semantic} (\emph{dense}) retrieval \cite{yang2020multilingual} and \emph{lexical} (\emph{sparse}) retrieval \cite{robertson2009probabilistic}. \emph{Semantic retrieval} maps both queries and documents into a 
shared high-dimensional dense vector space using a semantic embedding strategy such as Sentence-BERT \cite{reimers2019sentence}. It measures and indexes the \emph{similarity} between a query $q$ and a text chunk $c_i$ ($i=1,2,\cdots$), e.g., by cosine similarity: 
\begin{align} \label{eq:sim}
\mathrm{sim}(q, c_i) = \frac{\mathbf{v}_q \cdot \mathbf{v}_{c_i}}{\|\mathbf{v}_q\| \, \|\mathbf{v}_{c_i}\|},    
\end{align}
where $\mathbf{v}_q$ and $\mathbf{v}_{c_i}$ denote the embedding vectors of the query and the text chunk, respectively. Each text chunk $c_i$ is obtained by splitting the corpus into segments of a fixed (or bounded) length. The retriever selects the top-$k$ chunks by similarity, capturing semantic relationships and enabling the system to identify relevant information even when the wording differs.

In contrast, \emph{lexical retrieval} matches documents to a query primarily through exact (or near-exact) term overlap. It represents the query and each document as sparse vectors over the vocabulary and ranks documents using token overlap with term-weighting schemes. For example, BM25 \cite{robertson2009probabilistic} is a widely used probabilistic ranking function that estimates a document’s relevance to a query based on term frequency, inverse document frequency, and document-length normalization. 
Given a user query $q=\{t_1,t_2,\ldots,t_m\}$, where $t_j$ denotes the $j$-th term of the query, and a document chunk $c_i$, the BM25 relevance score is computed as: 
\begin{multline} \label{eq:bm25}
\mathrm{BM25}(q, c_i) = \\
\sum_{j=1}^m\mathrm{IDF}(t_j) \cdot
\frac{f(t_j, c_i) \cdot (k_1 + 1)}
{f(t_j, c_i) + k_1 \cdot 
\left( 1 - b + b \cdot \frac{N}{\mathrm{avgdl}} \right)},
\end{multline}
where $f(t_j, c_i)$ denotes the frequency of term $t_j$ in $c_i$, $N$ represents the chunk length, and $\mathrm{avgdl}$ is the average document length in the corpus. 
The parameters $k_1$ and $b$ control the influence of term frequency and chunk length on the relevance score, respectively. The inverse document frequency term, $\mathrm{IDF}(t_j)$, reflects how rare $t_j$ is in the corpus, assigning higher weight to terms that appear in fewer documents.
Since BM25 emphasizes lexical overlap, it is well suited to queries involving precise terminology and structured identifiers.

\subsubsection{Motivation}

Power system analysis often requires retrieving relevant information from large-scale power grid datasets (e.g., load and generation profiles) and from domain-specific documents (e.g., planning and operation guides, industry standards, and regulations). Conventional RAG approaches are effective for document-centric retrieval, but they are less reliable for retrieving information from large-scale power grid datasets. This limitation arises because power system data are highly structured and organized according to explicit schemas (e.g., buses, lines, generators, loads, and transformers). The meaning of a record often depends on its fields and its relationships to other components. Representing such structured data as unstructured text chunks can obscure these dependencies and reduce retrieval accuracy, particularly at large scales.
Moreover, the information needed to answer a query is often scattered across multiple components of the grid rather than contained within a small set of contiguous passages. As a result, retrieving only based on semantic similarity may provide insufficient coverage for tasks that require integrating numerical information across various components, which makes accurate and complete responses harder to obtain.

For example, consider the query on a large grid case: 
\begin{query} \label{query:ex2}
   ``\textit{Run AC power flow on the Texas 2k-bus grid\footnote{
The Texas 2k-bus grid is a synthetic  power grid dataset \cite{tamu_electricgrids} developed by the Texas A\&M team, which is added to X-GridAgent for large-scale testing.} and list the top 30 power lines with the highest loading ratios}."
\end{query}

\begin{figure}
	\centering
    \includegraphics[width=0.49\textwidth]{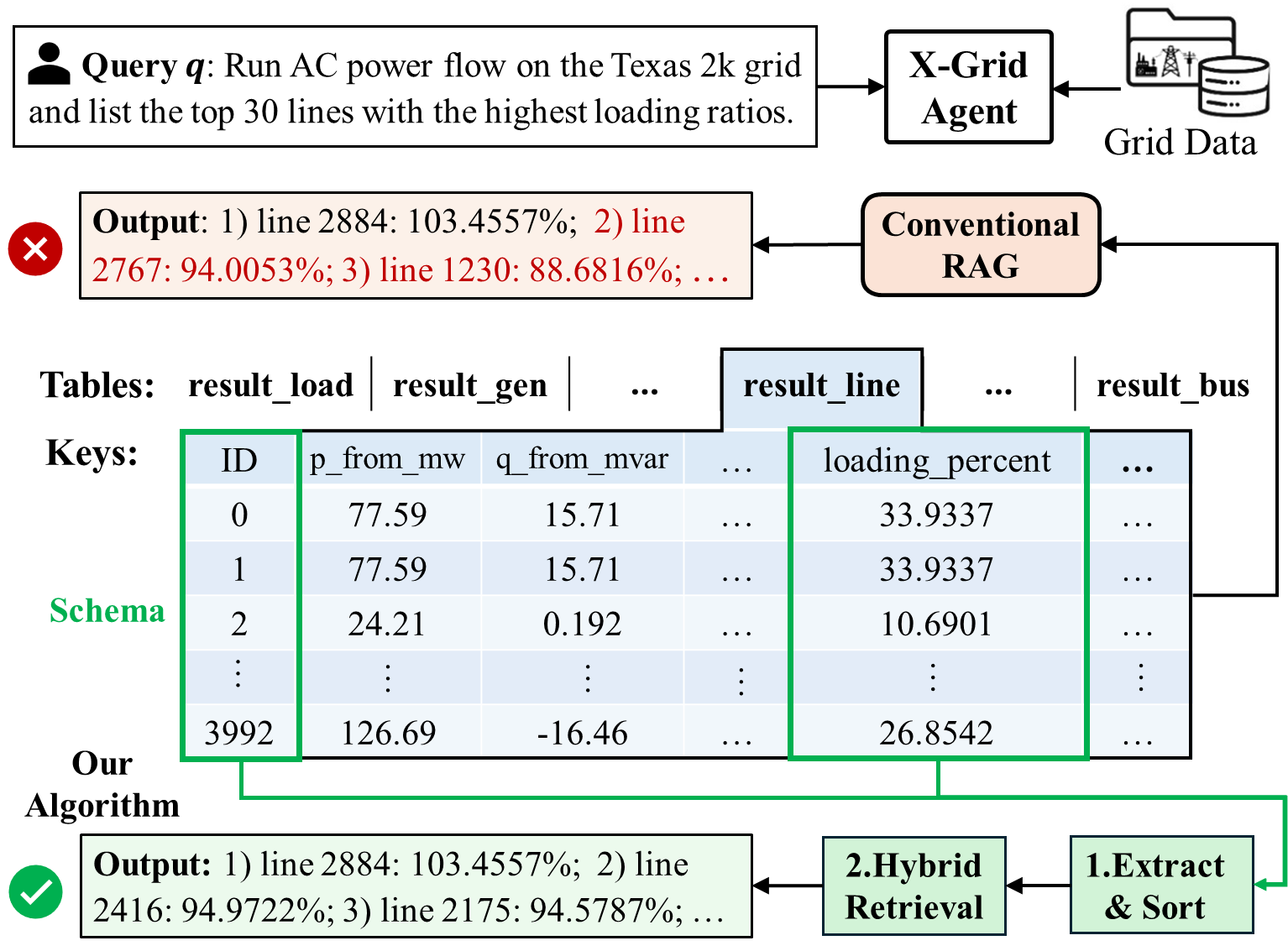}
    \vspace{-10pt}
	\caption{Comparison between conventional RAG methods and the proposed schema-adaptive hybrid RAG algorithm.
	}
	\label{fig:rag}
\end{figure}

To answer this query, X-GridAgent first calls the function \emph{solve\_AC\_powerflow()} in the \texttt{PowerFlow} server to execute the power flow calculation. It produces many result tables,  such as \texttt{result\_load}, \texttt{result\_gen}, \texttt{result\_line}, \texttt{result\_bus}, etc.
A conventional RAG method may retrieve the \texttt{result\_line} table as it has the most relevant column [loading\_percent]. However, \texttt{result\_line} is a large structured table with 3,992 rows (one per transmission line) and 14 feature columns, including [{ID}], [{from\_bus}], [{to\_bus}], [{p\_from\_mw}], [{q\_from\_mvar}], [loading\_percent], [{pl\_mw}], and more, as shown in Figure~\ref{fig:rag}. Conventional RAG methods often flatten this table into unstructured text and thus fail to retrieve the right lines with the highest loading percentages.

\subsubsection{Our Algorithm} 

To address these limitations of conventional RAG methods, we propose a novel algorithm called \emph{schema-adaptive hybrid RAG} for effectively retrieving information from large-scale power grid datasets. It aligns retrieval with both the physical structure and the semantic schema of power grid data. 
Specifically, the proposed algorithm consists of two steps: (1) \emph{schema-adaptive selection} and (2) \emph{hybrid retrieval}. In the first step, an LLM-based selection agent $\sS$ is constructed to generate a structured schema $\sX_q$ based on the original datasets $\sD_{\mathrm{ori}}$ and tailored to the retrieval query $q$:
\begin{align}
\mathcal{X}_q = \sS(\sD_\mathrm{ori}, q).
\end{align}
Here, the schema $\sX_q$ specifies the tables and columns relevant to fulfilling the retrieval query. For example, for the case shown in Figure \ref{fig:rag}, the selection agent $\sS$ outputs the schema: 
\begin{align}
\mathcal{X}_q &\coloneqq \Big\{ \text{``}\texttt{keyword}\text{"}:[\text{loading\_percent}], \ \text{``}\texttt{Table}\text{"}:  \nonumber \\ &\big\{\text{``}\texttt{res\_line}\text{"}: [\text{``ID"}, \text{``loading\_percent"}]\big\} \Big\},
\end{align}
which selects the keywords and identifies the corresponding tables and feature columns from the original dataset. 
Then, a reduced dataset is constructed based on the selected tables and columns. Moreover, given that power grid analysis typically focuses on extreme values (e.g., highest line loading ratios or lowest bus voltages), a sorting algorithm is applied to rank all rows within each table by the magnitude of the corresponding keyword column values. This results in a refined dataset $\sD_{\mathrm{ref}}$ used for the subsequent information retrieval.

In the second step, we construct a hybrid retriever that combines semantic retrieval with lexical retrieval (BM25) to leverage their complementary strengths. The hybrid relevance score between the query $q$ and each chunk $c_i$  of the refined dataset $\sD_{\mathrm{ref}}$ is computed as \eqref{eq:hyb}:
\begin{align} \label{eq:hyb}
s_{\text{hyb}}(q, c_i) = \lambda \cdot \mathrm{BM25}(q, c_i) + (1 - \lambda) \cdot \mathrm{sim}(q, c_i),    
\end{align}
which is a linear combination of \eqref{eq:sim} and \eqref{eq:bm25}, and the weight \(\lambda \in [0,1]\) controls the trade-off between semantic and lexical retrieval.  The hybrid retriever then selects the top-\(k\) chunks 
with the highest values of $s_{\text{hyb}}(q, c_i)$
for query $q$. 

In this way, our proposed \emph{schema-adaptive hybrid RAG} algorithm reduces both the volume and complexity of power grid data by extracting query-relevant information, and enhances overall retrieval efficiency, relevance, and accuracy through a hybrid approach that integrates semantic and lexical retrieval. As shown in Figure \ref{fig:rag}, our algorithm successfully retrieves the correct line information in response to Query \ref{query:ex2}.

%% file: Experiment.tex
\section{Implementation and Experiments} \label{sec:exper}

In this section, we introduce the implementation of the X-GridAgent system, outline the test experiments, and present its performance across various user queries.

\begin{figure}
	\centering
	\includegraphics[width=0.488\textwidth]{visua.png}
    \vspace{-10pt}
	\caption{The user interface of the X-GridAgent system. (A user enters a query in the chat window and clicks “SEND” to submit it. X-GridAgent performs reasoning and analysis, then displays the generated plan and execution results. Clicking “CLEAR” clears the historical memory and starts a new chat. The figure shows the system’s response to a user query requesting a visualization of the Texas 2k-bus grid network, and a follow-up query to run a DC optimal power flow (OPF) analysis is currently being typed.) } 
	\label{fig:interface}
\end{figure}

\begin{table*}
\centering
\caption{Experiments and test results of X-GridAgent across various user queries}
\begin{tabular}{@{}L{7cm}L{2.3cm}L{7.5cm}@{}}
\toprule
 \textbf{User Query} & \textbf{Server Use}& \textbf{X-GridAgent Output} (some details are omitted for brevity) \\
\midrule
 \emph{Q1 (Grid dataset retrieval)}. ``Load the IEEE-39 bus system. What are the connected buses for line 11? What are the voltage limits of bus 3?" & Task 1: 
 \texttt{Retrieval}  & - Connected buses for line index 11: from\_bus=5, to\_bus=10
 
- Voltage limits for bus 3: vm\_min=0.94 pu, vm\_max=1.06 pu \\
\midrule
\emph{Q2 (Documentation retrieval)}. ``In ERCOT operations, what is the maximum delay allowed for under-frequency relays serving load resources?" & Task 1: \texttt{Retrieval} & The under-frequency relay must have a delay of no more than 20 cycles. *Citation: ERCOT Nodal Operating Guides, Section 2.3.1.2, paragraph (6)(a) $\cdots$ \\
\midrule
\emph{Q3 (Topology)}. ``For the IEEE 39-bus system, what is the shortest path between bus 1 and bus 10?" & Task 1: \texttt{Topology} & - path\_bus\_indices: [1, 2, 3, 4, 5, 10]

- path\_edges: [[1, 2], [2, 3], [3, 4], [4, 5], [5, 10]] \\
\midrule
 \emph{Q4 (Power flow analysis)}. ``Run AC power flow on the IEEE 39-bus system using the fast-decoupled method. Show me the convergence result, the total active power loss, and the active power flow value of line 11."
 & Task 1: \texttt{PowerFlow} & 
- Converged: `True'

- Total active power loss: 43.6423 MW

- Line index 11 active power flow: 
 line 11 p\_from\_mw: -322.6527 MW, 
 line 11 p\_to\_mw: 323.3765 MW \\
 \midrule
  \emph{Q5 (Optimal Power Flow)}. ``Run DC OPF on the Texas 2k grid and return the optimal generation profile and total cost."  
 & Task 1: \texttt{OPF}  & - Total generation cost (OPF objective): 184290.887 
 
- Generation profile (p\_mw): gen\_id 260: 1100; gen\_id 158: 1100; gen\_id 159: 1100; gen\_id 412: 600; gen\_id 428: 525; $\cdots$ 
\\
  \midrule
   \emph{Q6 (Contingency)}. For the IEEE 118-bus system, perform an N-1 contingency analysis to evaluate the impact of disconnecting line 171. Summarize the limits violation.
   
 & Task 1: \texttt{Contingency}  & 
Post-contingency limit violations:

- Bus voltage (min Vm) violation: bus 117 has Vm = 0.929 pu, below Vmin = 0.94 pu (undervoltage of 0.011 pu).

- Thermal overloads: none reported.
     \\
     \midrule
   \emph{Q7 (Visualization)}.  
 ``For the Texas 2k grid, show me the network topology." 
 & Task 1: \texttt{Plot}  & (See Figure \ref{fig:interface} for the output and visualization of the Texas grid.) \\
   \midrule
   \emph{Q8 (Short-circuit analysis)}.  
 ``For the IEEE 300-bus system, suppose that there is a three-phase fault at bus 35, what are the maximal short-circuit currents? Summarize the results." 
 & Task 1: \texttt{ShortCircuit}  &
Maximal short-circuit (SC) results at Bus 35: 

$\vert$ Quantity $\vert$ Value $\vert$ Unit $\vert$

$\vert$ Ikss (steady-state SC current) $\vert$ 2.5719 $\vert$ kA $\vert$

$\vert$ Skss (SC apparent power) $\vert$ 1536.84 $\vert$ MVA  $\vert$ $\ \cdots$



 \\
    \midrule
 \emph{Q9 (Complex multi-step query)}. ``Can you run AC power flow on the IEEE 118-bus grid to find the top 5 most heavily loaded lines, then run contingency analysis on those and summarize any operational limit violations?" 
 & Task 1: \texttt{PowerFlow} $\to$ Task 2: \texttt{Retrieval}  $\to$ Task 3: \texttt{Contingency}   & 
- Line 6 outage: 1 voltage violation (bus 37 at 0.939 pu $<$ 0.94 pu); no branch overloads.

- Line 7 outage: 1 voltage violation (bus 37 at 0.939 pu $<$  0.94 pu); no branch overloads.

- Line 34 outage: no operational limit violations.

- Line 128 outage: no operational limit violations.

- Line 90 outage: no operational limit violations.
\\
     \midrule
 \emph{Q10 (Complex multi-step query)}. ``For the IEEE 39-bus system, consider three scenarios: Scenario 1 is the base case, Scenario 2 takes line 11 out of service, and Scenario 3 doubles all loads. Run OPF for each scenario, and compare and summarize the total generation cost and the generation outputs of generators 1, 2, and 3 across the scenarios." 
 & Task 1: \texttt{OPF} $\to$ Task 2: \texttt{Edit} $\to$  Task 3: \texttt{OPF}$\to$ 
 Task 4: \texttt{Edit} $\to$  Task 5: \texttt{OPF} $\to$ Task 6: \texttt{Retrieval} &Concise comparison:

- Scenario 2 vs Scenario 1: total generation cost increased from 41872.30 to 42115.34 ($\Delta$ $\approx$ +243.04). Generator dispatch shifted: Gen 1 decreased by $\sim$100.18 MW, Gen 2 decreased by $\sim$12.55 MW, while Gen 3 stayed unchanged ($\sim$508 MW).  

- Scenario 3: the AC OPF did not converge under doubled loads, so no reliable cost or dispatch values are available.\\
     \midrule
 \emph{Q11 (Complex multi-step query)}. ``From the ERCOT Operating Guide, what are the normal operational voltage limits for buses with a nominal voltage of 138 kV? Convert these voltage limits to per-unit values based on the nominal voltage, set those as Vmin/Vmax for all buses in the IEEE 118-bus system, then run an OPF and report  total generation cost and the resulting generation profile." & Task 1: \texttt{Retrieval} $\to$ Task 2: \texttt{Edit} $\to$  Task 3: \texttt{OPF}  & - Min voltage magnitude: 131.1 kV (0.95 pu on 138 kV base)
 
- Max voltage magnitude: 144.9 kV (1.05 pu on 138 kV base)

- AC OPF (case118, with bus voltage limits set to 0.95-1.05 pu):
 
*Converged: true.  *Total generation cost: 129746.0795961705

*Full generation dispatch profile (p\_mw, q\_mvar) 

$\vert$ gen id $\vert$  p\_mw $\vert$  q\_mvar $\vert$ 

$\vert$  0 $\vert$  26.92149967760338 $\vert$  14.999962115216187 $\vert$ 

$\vert$  1 $\vert$  1.6543434098907168e-05 $\vert$  70.34323261531028 $\vert$  $\ \cdots$ \\
%
\bottomrule
\end{tabular}
\label{tab:experi}
\end{table*}

\subsection{X-GridAgent Setup and Data Sources}

For the X-GridAgent system, we employed the OpenAI GPT-5 API 
as the LLM, serving as the core reasoning and language-understanding engine for interpreting and responding to user queries, while other LLM APIs can also be used. To enable professional power system analysis, we integrated the open-source Python-based software toolbox \emph{Pandapower} \cite{pandapower2018}, which supports power system modeling, analysis, and optimization. A number of standard power network cases (e.g., the IEEE 39-bus and 118-bus systems) \cite{pandapower_networks} and large-scale synthetic grid datasets (e.g., the Texas 2k grid) \cite{tamu_electricgrids} have been incorporated into X-GridAgent for testing purposes. 
 To enhance domain-specific expertise and ensure regulatory compliance, we embedded many publicly available documents (e.g., the ERCOT Planning Guide \cite{ercot_planning_guides} 
 and ERCOT Nodal Operating Guides \cite{ercot_nodal_operating}) into a documentation database, from which one can retrieve relevant information when responding to grid-related queries.
Moreover, we developed a user-friendly interface using PySide6 \cite{pyside6}, as shown in Figure \ref{fig:interface}, enabling users to interact with the X-GridAgent system solely through natural-language conversations. 
We showcase X-GridAgent in action in a demonstration video available online  \cite{gridagent_demo_video}.

\subsection{Testing Experiments}

We evaluate the X-GridAgent system using a diverse set of power system analysis queries. Table \ref{tab:experi} presents 11 representative test cases along with their outputs. Queries 1-8 are relatively simple and are used to test the basic functionality of X-GridAgent, while Queries 9-11 are more complex and require multi-step reasoning and coordination. 
All outputs generated by X-GridAgent have been verified to be correct by comparing them against solutions manually obtained through conventional methods. To assess the reliability of the system, we executed each query 30 times, and the success rate was 100\%, indicating consistent performance. We set the \texttt{temperature} parameter of the LLM to zero to mitigate stochastic variability in the model’s responses. While the exact wording of the outputs varies slightly between runs, the substantive answers remain accurate and consistent, demonstrating the reliability of the X-GridAgent system. In addition, the scalability of X-GridAgent is empirically verified and stems from its reliance on professional power system analysis tools for all domain-specific computations, allowing it to scale up to large systems (e.g., the Texas 2k-bus grid) and real-world power grids.

%% file: Conclusion.tex
\section{Conclusion}\label{sec:conclusion}

In this paper, we introduced X-GridAgent, a novel LLM-powered agentic AI system designed to automate comprehensive power grid analysis through natural language interaction. X-GridAgent demonstrates strong capabilities in interpreting user queries, dynamically generating task-specific workflows, and interfacing with domain-specific tools and databases to deliver trustworthy and interpretable results across a broad range of power system analysis tasks. 
Its three-layer hierarchical architecture offers flexibility and extensibility, allowing the seamless integration of new tools, data sources, and analytical capabilities. The current version, regarded as X-GridAgent 1.0, provides a foundation for extensive future enhancements. For example, as the present system focuses on steady-state power grid analysis, future work  will aim to support dynamic and transient grid studies, as well as advanced decision-making functionalities. In addition, the system architecture can be further improved to enhance workflow stability and reliability.

%% file: Reference.tex
\bibliography{IEEEabrv, ref}

%% file: main.bbl
\begin{thebibliography}{10}
\providecommand{\url}[1]{#1}
\csname url@samestyle\endcsname
\providecommand{\newblock}{\relax}
\providecommand{\bibinfo}[2]{#2}
\providecommand{\BIBentrySTDinterwordspacing}{\spaceskip=0pt\relax}
\providecommand{\BIBentryALTinterwordstretchfactor}{4}
\providecommand{\BIBentryALTinterwordspacing}{\spaceskip=\fontdimen2\font plus
\BIBentryALTinterwordstretchfactor\fontdimen3\font minus \fontdimen4\font\relax}
\providecommand{\BIBforeignlanguage}[2]{{%
\expandafter\ifx\csname l@#1\endcsname\relax
\typeout{** WARNING: IEEEtran.bst: No hyphenation pattern has been}%
\typeout{** loaded for the language `#1'. Using the pattern for}%
\typeout{** the default language instead.}%
\else
\language=\csname l@#1\endcsname
\fi
#2}}
\providecommand{\BIBdecl}{\relax}
\BIBdecl

\bibitem{muhtadi2021distributed}
A.~Muhtadi, D.~Pandit, N.~Nguyen, and J.~Mitra, ``Distributed energy resources based microgrid: Review of architecture, control, and reliability,'' \emph{IEEE Transactions on Industry Applications}, vol.~57, no.~3, pp. 2223--2235, 2021.

\bibitem{chen2025electricity}
X.~Chen, X.~Wang, A.~Colacelli, M.~Lee, and L.~Xie, ``Electricity demand and grid impacts of {AI} data centers: Challenges and prospects,'' \emph{arXiv preprint arXiv:2509.07218}, 2025.

\bibitem{chen2024towards}
X.~Chen, H.~Chao, W.~Shi, and N.~Li, ``Towards carbon-free electricity: A flow-based framework for power grid carbon accounting and decarbonization,'' \emph{Energy Conversion and Economics}, vol.~5, no.~6, pp. 396--418, 2024.

\bibitem{11131348}
Q.~Zhang and L.~Xie, ``{PowerAgent}: A road map toward agentic intelligence in power systems: Foundation model, model context protocol, and workflow,'' \emph{IEEE Power and Energy Magazine}, vol.~23, no.~5, pp. 93--101, 2025.

\bibitem{chen2022reinforcement}
X.~Chen, G.~Qu, Y.~Tang, S.~Low, and N.~Li, ``Reinforcement learning for selective key applications in power systems: Recent advances and future challenges,'' \emph{IEEE Transactions on Smart Grid}, vol.~13, no.~4, pp. 2935--2958, 2022.

\bibitem{zhao2023survey}
W.~X. Zhao, K.~Zhou, J.~Li, T.~Tang, X.~Wang, Y.~Hou, Y.~Min, B.~Zhang, J.~Zhang, Z.~Dong \emph{et~al.}, ``A survey of large language models,'' \emph{arXiv preprint arXiv:2303.18223}, vol.~1, no.~2, 2023.

\bibitem{achiam2023gpt}
J.~Achiam, S.~Adler, S.~Agarwal, L.~Ahmad, I.~Akkaya, F.~L. Aleman, D.~Almeida, J.~Altenschmidt, S.~Altman, S.~Anadkat \emph{et~al.}, ``Gpt-4 technical report,'' \emph{arXiv preprint arXiv:2303.08774}, 2023.

\bibitem{team2023gemini}
G.~Team, R.~Anil, S.~Borgeaud, J.-B. Alayrac, J.~Yu, R.~Soricut, J.~Schalkwyk, A.~M. Dai, A.~Hauth, K.~Millican \emph{et~al.}, ``Gemini: a family of highly capable multimodal models,'' \emph{arXiv preprint arXiv:2312.11805}, 2023.

\bibitem{zeng2023large}
F.~Zeng, W.~Gan, Y.~Wang, N.~Liu, and P.~S. Yu, ``Large language models for robotics: A survey,'' \emph{arXiv preprint arXiv:2311.07226}, 2023.

\bibitem{hadi2023large}
M.~U. Hadi, R.~Qureshi, A.~Shah, M.~Irfan, A.~Zafar, M.~B. Shaikh, N.~Akhtar, J.~Wu, S.~Mirjalili \emph{et~al.}, ``Large language models: a comprehensive survey of its applications, challenges, limitations, and future prospects,'' \emph{Authorea preprints}, vol.~1, no.~3, pp. 1--26, 2023.

\bibitem{11086353}
M.~Jia, Z.~Cui, and G.~Hug, ``Enhancing {LLMs} for power system simulations: A feedback-driven multi-agent framework,'' \emph{IEEE Transactions on Smart Grid}, vol.~16, no.~6, pp. 5556--5572, 2025.

\bibitem{deng2025power}
K.~Deng, Y.~Zhou, H.~Zeng, Z.~Wang, and Q.~Guo, ``Power grid model generation based on the tool-augmented large language model,'' \emph{IEEE Transactions on Power Systems}, 2025.

\bibitem{10747635}
S.~Jin and S.~Abhyankar, ``{ChatGrid}: Power grid visualization empowered by a large language model,'' in \emph{2024 IEEE Workshop on Energy Data Visualization (EnergyVis)}, 2024, pp. 12--17.

\bibitem{11129042}
F.~Amjad, T.~Korõtko, and A.~Rosin, ``Review of {LLMs} applications in electrical power and energy systems,'' \emph{IEEE Access}, vol.~13, pp. 150\,951--150\,969, 2025.

\bibitem{majumder2024exploring}
S.~Majumder, L.~Dong, F.~Doudi, Y.~Cai, C.~Tian, D.~Kalathil, K.~Ding, A.~A. Thatte, N.~Li, and L.~Xie, ``Exploring the capabilities and limitations of large language models in the electric energy sector,'' \emph{Joule}, vol.~8, no.~6, pp. 1544--1549, 2024.

\bibitem{durante2024agent}
Z.~Durante, Q.~Huang, N.~Wake, R.~Gong, J.~S. Park, B.~Sarkar, R.~Taori, Y.~Noda, D.~Terzopoulos, Y.~Choi \emph{et~al.}, ``Agent {AI}: Surveying the horizons of multimodal interaction,'' \emph{arXiv preprint arXiv:2401.03568}, 2024.

\bibitem{10849561}
D.~B. Acharya, K.~Kuppan, and B.~Divya, ``Agentic {AI}: Autonomous intelligence for complex goals—a comprehensive survey,'' \emph{IEEE Access}, vol.~13, pp. 18\,912--18\,936, 2025.

\bibitem{fan2024survey}
W.~Fan, Y.~Ding, L.~Ning, S.~Wang, H.~Li, D.~Yin, T.-S. Chua, and Q.~Li, ``A survey on {RAG} meeting {LLMs}: Towards retrieval-augmented large language models,'' in \emph{Proceedings of the 30th ACM SIGKDD conference on knowledge discovery and data mining}, 2024, pp. 6491--6501.

\bibitem{singh2025agentic}
A.~Singh, A.~Ehtesham, S.~Kumar, and T.~T. Khoei, ``Agentic retrieval-augmented generation: A survey on agentic {RAG},'' \emph{arXiv preprint arXiv:2501.09136}, 2025.

\bibitem{anthropic2025mcp}
Anthropic, ``Introducing the model context protocol,'' [Online]. Available: \url{https://www.anthropic.com/news/model-context-protocol}, 2025, accessed: 2025-10-20.

\bibitem{hou2025model}
X.~Hou, Y.~Zhao, S.~Wang, and H.~Wang, ``Model context protocol ({MCP}): Landscape, security threats, and future research directions,'' \emph{arXiv preprint arXiv:2503.23278}, 2025.

\bibitem{11070004}
R.~Zou, X.~Zhou, Y.~Cheng, W.~Liu, X.~Wang, J.~Zhao, and X.~Cai, ``A large language model-based agent for automated bidding strategy generation in electricity markets,'' in \emph{2025 IEEE International Conference on Power and Integrated Energy Systems (ICPIES)}, 2025, pp. 475--480.

\bibitem{ren2025can}
X.~Ren, C.~S. Lai, G.~Taylor, and Z.~Guo, ``Can large language model agents balance energy systems?'' \emph{arXiv preprint arXiv:2502.10557}, 2025.

\bibitem{yang2025large}
X.~Yang, C.~Lin, Y.~Yang, Q.~Wang, H.~Liu, H.~Hua, and W.~Wu, ``Large language model powered automated modeling and optimization of active distribution network dispatch problems,'' \emph{IEEE Transactions on Smart Grid}, 2025.

\bibitem{zhaolarge}
H.~Zhao, Y.~Cheng, D.~Xiang, X.~Zhou, J.~Zhao, X.~Cai, and Z.~Y. Dong, ``Large language model-based power dispatch agent: Framework, application, and challenges,'' \emph{Application, and Challenges}, 2025.

\bibitem{zhu2025powercon}
Y.~Zhu, Y.~Zhou, and W.~Wei, ``{PowerCon}: A collaborative framework integrating {LLM} and power {AI} for interactive and regulation-compliant day-ahead dispatch,'' \emph{Authorea Preprints}, 2025.

\bibitem{li2025optdispro}
Z.~Li, H.~Yang, Y.~Liu, Y.~Xiang, H.~Gao, J.~Liu, and J.~Liu, ``{OptDisPro}: {LLM}-based multi-agent framework for flexibly adapting heuristic optimal disflow,'' \emph{IEEE Transactions on Smart Grid}, 2025.

\bibitem{bernier2025powergraph}
F.~Bernier, J.~Cao, M.~Cordy, and S.~Ghamizi, ``{PowerGraph-LLM}: Novel power grid graph embedding and optimization with large language models,'' \emph{IEEE Transactions on Power Systems}, 2025.

\bibitem{zhang2025grid}
Y.~Zhang, A.~M. Saber, A.~Youssef, and D.~Kundur, ``{Grid-Agent}: An llm-powered multi-agent system for power grid control,'' \emph{arXiv preprint arXiv:2508.05702}, 2025.

\bibitem{saha2025dragent}
B.~K. Saha, V.~Aarthi, and O.~Naidu, ``{DrAgent}: An agentic approach to fault analysis in power grids using large language models,'' in \emph{2025 International Conference on Artificial Intelligence in Information and Communication (ICAIIC)}.\hskip 1em plus 0.5em minus 0.4em\relax IEEE, 2025, pp. 0938--0945.

\bibitem{jin2025gridmind}
H.~Jin, K.~Kim, and J.~Kwon, ``{GridMind}: {LLMs}-powered agents for power system analysis and operations,'' in \emph{Proceedings of the SC'25 Workshops of the International Conference for High Performance Computing, Networking, Storage and Analysis}, 2025, pp. 560--568.

\bibitem{badmus2025powerchain}
E.~O. Badmus, P.~Sang, D.~Stamoulis, and A.~Pandey, ``{PowerChain}: A verifiable agentic {AI} system for automating distribution grid analyses,'' \emph{arXiv preprint arXiv:2508.17094}, 2025.

\bibitem{gridagent_demo_video}
X.~Chen and Y.~Wen, ``{X-GridAgent} demonstration video,'' [Online]. Available: \url{https://www.youtube.com/watch?v=JUqpwO6NncY}, 2025, accessed: 2025-12-17.

\bibitem{pandapower2018}
L.~Thurner, A.~Scheidler, F.~Sch{\"a}fer, J.~Menke, J.~Dollichon, F.~Meier, S.~Meinecke, and M.~Braun, ``pandapower — an open-source python tool for convenient modeling, analysis, and optimization of electric power systems,'' \emph{IEEE Transactions on Power Systems}, vol.~33, no.~6, pp. 6510--6521, Nov 2018.

\bibitem{DIN_EN_60909_0}
\emph{Short-circuit currents in three-phase a.c. systems -- Part 0: Calculation of currents}, DIN; IEC Std. DIN EN 60\,909-0 (IEC 60\,909-0), 2016, german version EN 60909-0:2016.

\bibitem{bo2024reflective}
X.~Bo, Z.~Zhang, Q.~Dai, X.~Feng, L.~Wang, R.~Li, X.~Chen, and J.-R. Wen, ``Reflective multi-agent collaboration based on large language models,'' \emph{Advances in Neural Information Processing Systems}, vol.~37, pp. 138\,595--138\,631, 2024.

\bibitem{marvin2023prompt}
G.~Marvin, N.~Hellen, D.~Jjingo, and J.~Nakatumba-Nabende, ``Prompt engineering in large language models,'' in \emph{International conference on data intelligence and cognitive informatics}.\hskip 1em plus 0.5em minus 0.4em\relax Springer, 2023, pp. 387--402.

\bibitem{zhao2024dense}
W.~X. Zhao, J.~Liu, R.~Ren, and J.-R. Wen, ``Dense text retrieval based on pretrained language models: A survey,'' \emph{ACM Transactions on Information Systems}, vol.~42, no.~4, pp. 1--60, 2024.

\bibitem{ma2023query}
X.~Ma, Y.~Gong, P.~He, N.~Duan \emph{et~al.}, ``Query rewriting in retrieval-augmented large language models,'' in \emph{The 2023 Conference on Empirical Methods in Natural Language Processing}, 2023.

\bibitem{yang2020multilingual}
Y.~Yang, D.~Cer, A.~Ahmad, M.~Guo, J.~Law, N.~Constant, G.~H. Abrego, S.~Yuan, C.~Tar, Y.-H. Sung \emph{et~al.}, ``Multilingual universal sentence encoder for semantic retrieval,'' in \emph{Proceedings of the 58th Annual Meeting of the Association for Computational Linguistics: System Demonstrations}, 2020, pp. 87--94.

\bibitem{robertson2009probabilistic}
S.~Robertson, H.~Zaragoza \emph{et~al.}, ``The probabilistic relevance framework: Bm25 and beyond,'' \emph{Foundations and Trends{\textregistered} in Information Retrieval}, vol.~3, no.~4, pp. 333--389, 2009.

\bibitem{reimers2019sentence}
N.~Reimers and I.~Gurevych, ``{Sentence-BERT}: Sentence embeddings using siamese {BERT}-networks,'' in \emph{Proceedings of the 2019 Conference on Empirical Methods in Natural Language Processing and the 9th International Joint Conference on Natural Language Processing (EMNLP-IJCNLP)}, 2019, pp. 3982--3992.

\bibitem{tamu_electricgrids}
``Electric grid test case repository,'' \url{https://electricgrids.engr.tamu.edu/}, 2025, accessed: 2025-12-13; synthetic electric grid datasets from Texas A\&M University.

\bibitem{pandapower_networks}
P.~Developers, ``Pandapower network examples,'' [Online]. Available: \url{https://pandapower.readthedocs.io/en/latest/networks.html}, 2025, accessed: 2025-12-17.

\bibitem{ercot_planning_guides}
{Electric Reliability Council of Texas (ERCOT)}, ``Current planning guides,'' [Online]. Available: \url{https://www.ercot.com/mktrules/guides/planning/current}, 2025, accessed: 2025-12-17.

\bibitem{ercot_nodal_operating}
------, ``Current nodal operating guides,'' [Online]. Available: \url{https://www.ercot.com/mktrules/guides/noperating/current}, 2025, accessed: 2025-12-17.

\bibitem{pyside6}
{The Qt Company}, ``{Qt for Python (PySide6)},'' \url{https://doc.qt.io/qtforpython/}, 2024, accessed: 2025-12.

\end{thebibliography}
